# A Systematic Literature Review on the Use of Machine Learning in Software Engineering

Nyaga Fred  m156796@edu.misis.ru,

Temkin, I.O.

National University of Science and Technology "MISiS", Moscow, 119049, Russia

## Abstract

Software engineering (SE) is a dynamic field that involves multiple phases all of which are necessary to develop sustainable software systems. Machine learning (ML), a branch of artificial intelligence (AI), has drawn a lot of attention in recent years thanks to its ability to analyze massive volumes of data and extract useful patterns from data. Several studies have focused on examining, categorising, and assessing the application of ML in SE processes. We conducted a literature review on primary studies to address this gap. The study was carried out following the objective and the research questions to explore the current state of the art in applying machine learning techniques in software engineering processes. The review identifies the key areas within software engineering where ML has been applied, including software quality assurance, software maintenance, software comprehension, and software documentation. It also highlights the specific ML techniques that have been leveraged in these domains, such as supervised learning, unsupervised learning, and deep learning.

# SECTION 1

**Introduction**

Software engineering (SE) is a rapidly changing field that involves designing, developing, maintaining, testing and evolving software systems in a systematic and controlled manner. Machine learning (ML), a subfield of artificial intelligence (AI), has received substantial attention in recent years due to its potential to analyse enormous volumes of data. This has led to the exploration of its potential applications in software engineering, including but not limited to defect detection, code quality assessment, requirement analysis, and software project management. In today's SE landscape, the demand for high-quality and maintainable source code is paramount. Software applications are becoming increasingly complex, developers are facing challenges in writing efficient and error-free code. Recent research shows that ML techniques can be seamlessly integrated into the software development process offering solutions to these challenges Pradel M. et al. (2018.

Advances in deep learning (DL) and natural language processing (NLP) have laid the foundation for the usage of machine learning tools and techniques to perform an extensive range of programming tasks. DL, a subset of ML, has revolutionised several industries such as language processing, image processing, text translation and generation Beese D. et al. (2023). In the context of SE, natural language processing (NLP) can be used to perform text-related tasks including software documentation, bug finding, code completion, and code translation, greatly boosting developer productivity and code quality Pauzi et al. (2023). Conversely, image processing techniques can be utilised for activities such as visual testing, image recognition, user interface design, and enhancing the user experience within software systems Chen J.S. (2022). The integration of these technologies not only streamlines software development processes but also improves the functionality and usability of software applications.

Programmers have long depended on traditional approaches (such as compilers) to analyse code; these techniques use accuracy and logical reasoning to understand and manipulate software Nielson F. et al. (2015). However, as software systems become more sophisticated, such approaches become less effective in solving particular issues. Existing software in SE must be updated and modified on a continuous basis as user needs, business logic, legislation, and technology evolve; software systems have to change and adapt in order to remain usable and effective Nielson F. et al. (2015) Canfora G. et al. (2011).

This method of continuously modifying and improving source code is known as software evolution; it has been a topic of research since the 1970s, popularised by Lehman [14]. Researchers are currently looking into ways of addressing challenges presented by   fast growing source code and the constant need for novel approaches in SE by focusing on ML techniques. These efforts seek to improve software evolutionary processes and address the dynamic nature of software development Canfora G. et al. (2011) de Oliveira (2014).  ML models are trained using software engineering data sourced from repositories, historical code changes, documentation, and bug reports to tailor them for SE tasks. Analyzing repositories allows researchers to gain empirical insights into software development practices, aiding in the management, maintenance, and evolution of software systems. The integration of machine learning (ML) has significantly elevated its importance in the realm of software engineering (SE), with a burgeoning research landscape delving into the application of ML techniques across diverse SE tasks.

This paper aims to provide a comprehensive and analysis on the current research advancements in the field of machine learning and software engineering. Furthermore, the review seeks to uncover the research challenges encountered, highlight existing opportunities for advancement, and pinpoint any gaps in the current research landscape. By synthesizing these aspects, this review aims to contribute to a deeper understanding of the utilization of machine learning in software engineering and pave the way for future research directions and innovations in this dynamic field.

# SECTION 2

**Related work**

There is an increasing body of research on the application of machine learning techniques in software engineering. Several studies have looked into the current applications of machine learning in SE and how they affect development processes and outcomes. Finding and fixing software bugs is one of the most challenging and costly tasks in software development. Bugs may emanate from a variety of sources, such as errors in requirements gathering, shortcomings in design, coding errors, integration and deployment issues. To address these shortcomings, Michael Pradel et al. (2021) conducted an overview of the use of deep learning (DL) to analyse programmes using neural software analysis - an alternative to traditional tools. The study makes a compelling case for the use of neural networks in bug detection with DeepBugs, a deep learning-based tool. DeepBugs produces promising results when detecting potential issues using natural language information within code. However, there are serious concerns about its reliance on artificially generated bugs and the lack of additional validation steps.

In related studies, Tian et al. (2015) sought to predict software defects during code changes using deep learning models. Their primary focus was on offering immediate feedback to developers during code reviews and early development stages to identify source code bugs and vulnerabilities. Conversely, Allamanis et al. (2021) concentrated on automatically fixing build errors using a novel neural network architecture named Graph2Diff. This study addressed the challenges associated with post-commit build errors by employing graph neural networks (GNNs) to model and rectify code changes.

*Software maintenance*. It is important to ensure that applications and software systems remain effective and relevant in the face of changing requirements and technological advancements. Several studies have looked into the application of machine learning in software maintenance. Predictive Maintenance and Defect Prediction. Hall et al. (2011) carried out a comprehensive literature review on fault prediction performance in software engineering, emphasising the effectiveness of decision trees, neural networks, and support vector machines (SVMs) in predicting software defects; these algorithms make use of historical defect data, code metrics, and other features to forecast flaws and facilitate proactive maintenance. Similarly, Tian et al. (2015) concentrated their study on the use of deep belief networks (DBNs) for just-in-time defect prediction. Their approach combined code complexity measures and historical defect data to deliver real-time feedback during the code review process. This integration sought to improve early defect detection and, as a result, reduce the number of software issues after release. In a comparable manner, Jindal et al. (2015) conducted a study to estimate software maintenance effort using neural networks. Their findings imply that neural network models can accurately predict the amount of maintenance required in software systems, thereby facilitating effective resource planning and allocation.

*Automated Bug Fixing and Localization*. Bug localisation is an essential aspect of software maintenance that identifies the exact location of issues in the codebase. Accurate and fast bug localization is critical for ensuring software quality since it allows developers to quickly identify and resolve issues. Several studies looked into alternative machine learning methods for improving bug localisation and automated bug fixing, resulting in more robust and maintainable software systems. For example, Yousofvand et al. (2023) proposed a hybrid approach that combines deep learning and model transformation through node classification. This method uses control flow graphs (CFGs) to convert source code into a format that deep learning models can easily examine. Similarly, Ma et al. proposed a new method centred around the utilization of Control Flow Graphs (CFGs) to capture the structural and logical flow of source code. Their approach (cFlow) leverages the inherent properties of CFGs to enhance the feature learning process, thus improving the accuracy of bug localization. This technique takes into account the flowing nature and intrinsic relationships between paths; cFlow includes a flow-based GRU designed for transferring semantics along execution paths; in addition, the programme structure is expressed using CFG, This structure is critical in capturing the complex relationships between different parts of the code, leading to better bug localization.

*Automated Code Generation and Optimization*. Automatic code generation utilizes AI models such as ML and deep learning (DL) to produce fragments of code or full programmes based on input requirements, specifications, and constraints. Several studies have looked into innovative methods for improving code generation. For example, Liao et al. (2010) published a seminal study on automated code generation, emphasizing formal and model-driven development methodologies for producing maintainable code. This study highlights MDD as a critical technique that transforms abstract models into executable code via a succession of refining processes. Perez et al. (2021) unlike Liao et al. (2010) explored the usage of pre-trained language models such as GPT-2 to generate code. This approach leverages the transformer architecture to capture complex patterns and relationships in text data, making it well-suited for code generation tasks. However, the study has several shortcomings that may limit its generalizability and applicability. For example, the datasets used are relatively small when compared to other code generation datasets focusing only on one programming language hence limiting the effectiveness of this approach. In a comparable study, Sebastián et al. (2020) proposed a novel approach that combines natural language processing (NLP) and machine learning techniques to generate source code for MVC-based applications.

*Code summarization and documentation*. Some studies have looked into effective approaches to document software; for example, Hashemi et al. (2020) developed a structured documentation framework aimed at ML projects. The findings of their study may help to improve the quality and usability of ML software documentation, benefiting both developers and users. However, the study is constrained because the paper does not describe the specific methodologies or tools utilised for data collecting, filtering, and classification of Stack Overflow Q&As. In contrast to Hashemi et al. (2020), Barone et al. (2017) concentrated solely on building a parallel corpus of Python functions and documentation texts. Their work contributes to automated code documentation and creation by creating a structured dataset for training machine learning algorithms. This dataset bridges the gap between code and natural language descriptions, allowing for more efficient automated documentation systems. The method uses sequence-to-sequence models to translate code into documentation and vice versa, illustrating machine learning's utility in software documentation. Additionally, McBurney and McMillan (2015) attempted to solve the problem of software documentation by developing techniques for automatic Java source code documentation. Their method combines static analysis with ML to generate summaries that provide context and understanding of the code, this approach could be beneficial when dealing with large codebases.

The literature review underlined the growing amount of research on the application of machine learning algorithms in different aspects of software engineering, such as bug identification, software maintenance, code generation, and documentation. Although these studies have made major contributions, there are still several gaps and limitations. One noteworthy shortcoming is the absence of comprehensive frameworks that combine different machine-learning techniques for rapid and efficient source code management. Most present systems focus on individual tasks, such as issue localization or defect prediction, but do not offer an exhaustive solution to the complex challenges of software maintenance.

Although several studies have investigated the application of deep learning for code analysis and generation, there has been little research into the incorporation of advanced deep learning architectures, such as transformer-based models and graph neural networks, for software engineering tasks. Furthermore, additional research is required on the actual use and integration of machine learning techniques into existing software development processes and tools. Many studies focus on theoretical methodologies or proof-of-concept implementations, however, there is little emphasis on how to effectively apply these strategies in real-world software engineering tasks.

# SECTION 3

## Methodology

This paper presents a systematic literature review to explore the use of machine learning techniques in software engineering. The methodology for this review is derived from established guidelines for conducting systematic literature reviews (Mahdi et al., 2021; Crawford et al., 2023). The review process begins with the identification of relevant research articles through a comprehensive search of electronic databases, followed by careful screening and selection of studies based on predefined inclusion and exclusion criteria. The selected studies are then analyzed and synthesized to extract key insights, trends, and challenges associated with the application of machine learning in software engineering.

## 3.1. Research Design

The systematic literature review (SLR) adheres to the criteria established by Kitchenham and Charters (2007) for performing systematic reviews in software engineering. The approach comprises three major phases: planning, execution, and reporting. Each phase includes several steps to ensure a thorough and unbiased evaluation of existing literature.

## 3.2. Planning the Review

Question synthesis is an important step in the development of a research review paper, particularly in a multidisciplinary field such as the intersection of machine learning and software engineering. This procedure ensures that the research is focused, thorough, and applicable. We used a well-defined approach to formulate our research questions (RQ) in accordance with the study's goals, which include investigating the applications, challenges, benefits, and future directions of ML in SE. We finally came up with the following RQs as the bases of our research:

**Machine Learning Uses in Software Engineering**

**RQ1:** What are the current applications of machine learning in software engineering, and how do they impact development processes and outcomes?

**RQ2:** How can machine learning enhance software maintenance, evaluation, and debugging efficiency and quality?

**Challenges of Integrating Machine Learning with Software Engineering**

**RQ3: What are the main challenges faced when integrating machine learning techniques into the software development workflow?**

**RQ4: How do issues like data quality, interpretability of models, and scalability affect the successful implementation of machine learning in software development**?

**Future Directions and Innovations in Machine Learning for Software Engineering**

**RQ5**: What are the emerging trends and future directions of machine learning in software engineering research and industry applications?

**RQ6**: How can advancements in machine learning algorithms and techniques drive innovation in software development processes and tools?

### 3.3. Search Strategy

Question synthesis is an important step in the development of a research review paper, particularly in a multidisciplinary field such as the intersection of machine learning and software engineering. This procedure ensures that the research is focused, thorough, and applicable. We used a well-defined approach to formulate our research questions (RQ) per the study's goals, which include investigating the applications, challenges, benefits, and future directions of ML in SE. We finally came up with the following RQs as the basis of our research:

**Machine Learning Uses in Software Engineering**

RQ1: What are the current applications of machine learning in software engineering, and how do they impact development processes and outcomes?

RQ2: How can machine learning enhance software maintenance, evaluation, and debugging efficiency and quality?

Challenges of Integrating Machine Learning **with Software Engineering**

**RQ3: What are the main challenges faced when integrating machine learning techniques into the software development workflow?**

**RQ4: How do issues like data quality, interpretability of models, and scalability affect the successful implementation of machine learning in software development?**

**Future Directions and Innovations in Machine Learning for Software Engineering**

RQ5: What are the emerging trends and future directions of machine learning in software engineering research and industry applications?

RQ6: How can advancements in machine learning algorithms and techniques drive innovation in software development processes and tools?

## 3.3. Search Strategy

A detailed search was undertaken across numerous electronic databases to find relevant studies. The search phrases were generated from the research questions and included keyword combinations such as "machine learning," "software engineering," "application," "algorithm," "dataset," "challenge," and "Deep Learning." The search approach entailed searching three main digital libraries: IEEE Xplore, ACM Digital Library, Google Scholar, arXiv.org and ScienceDirect.

## 3.4. Inclusion and Exclusion Criteria

Inclusion Criteria:

- studies published in peer-reviewed journals or conferences;

- studies that discuss the application of machine learning in any area of software engineering;

- studies published in English;

- studies that provide empirical evidence or case studies.

Exclusion Criteria:

- studies that are not directly related to software engineering;

- studies without a focus on machine learning applications;

- grey literature, such as technical reports, dissertations, and white papers;

- duplicate studies across different databases.

## 3.5. Conducting the review

The study selection process involved the following steps:

Initial Screening: The titles and abstracts of identified studies were examined to remove irrelevant papers.

*Full-Text Review*: Full texts of the remaining studies were assessed for eligibility based on the inclusion and exclusion criteria.

Quality Assessment: Selected studies were assessed for quality using a specified checklist based on Kitchenham and Charters (2007). The checklist includes requirements for conceptual clarity, methodology suitability, and results validity.

Figure 1 depicts the primary steps we took when deciding whether a study was relevant to the research topic or otherwise. The first stage required reading the abstract to gain an understanding of the study. Nonetheless, if the abstract was not straightforward, we moved on to the introduction and  full paper respectively  Van Vliet et al. (2008).

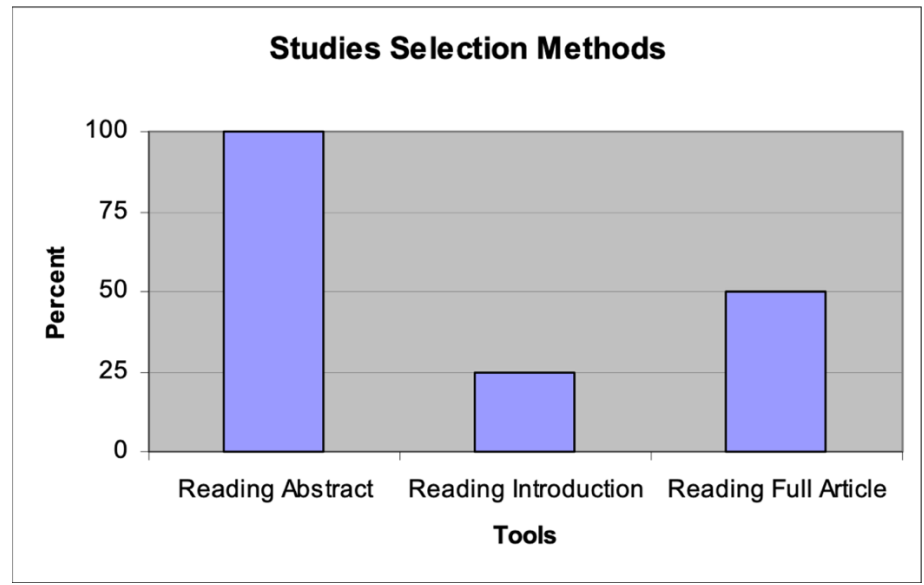

**Figure 1. Steps for Study Selection**

## 3.6. Reporting the Review

*Data Synthesis.*The gathered data was synthesised using a combination of qualitative and quantitative techniques. Descriptive statistical methods were used to summarise the distribution of research across various categories, including publication year, software engineering tasks, and machine learning algorithms. Thematic analysis was carried out to discover common themes, obstacles, and future directions mentioned in the research.

### 3.6.1 Presentation of Results

The SLR findings are provided in a systematic manner, in accordance with the research questions. Each section summarises the findings, supported by tables, figures, and direct quotes from the included research where applicable. Trends and gaps in the literature are emphasised, allowing for an extensive understanding of the present state of research on the application of machine learning in software engineering.

## SECTION 4

### Findings and Discussion

The literature review uncovered numerous uses of ML in SE tasks, some of which have a substantial impact on development processes and outcomes. The review was carried out in accordance with the research questions outlined in the methodology section. The results for each RQ are shown below:

**RQ1**: What are the current applications of machine learning in software engineering, and how do they impact development processes and outcomes?

Based on the reviewed literature, machine learning has been successfully applied to a variety of software engineering tasks, including:(Crawford et al., 2023)

- Defect prediction and code quality assessment (Mahdi et al., 2021)(Systematic Literature Review on the Use of Machine Learning in Software Engineering, n.d) - ML models can accurately predict bugs and defects in source code, enabling developers to focus testing and debugging efforts on the most problematic components.

- Software requirements engineering - ML techniques can be used to automatically extract, analyze, and prioritize software requirements from user stories, bug reports, and other natural language sources.

- Effort estimation and project management - ML models trained on historical project data can provide accurate estimates of development effort, timeline, and resource needs for new software projects.

ML has demonstrated significant benefits for these SE tasks, including improved efficiency, reduced costs, and enhanced product quality (Crawford et al., 2023)(Mahdi et al., 2021).

**RQ2**: How can machine learning enhance software maintenance, evaluation, and debugging efficiency and quality?

ML algorithms have shown promising results in several areas of software maintenance and quality assurance:

- Automated program repair - ML-based approaches can automatically generate bug fixes and patches, accelerating the maintenance process.

- Test case generation and prioritization - ML models can intelligently select and order test cases to maximize code coverage and defect detection.

- Automated code analysis and summarization - ML techniques like program embeddings can provide contextual summaries of code to aid developer comprehension and debugging.

**RQ3**: What are the main challenges faced when integrating machine learning techniques into the software development workflow?

Despite the benefits, several critical challenges hinder the successful integration of ML in software engineering(Liem & Panichella, 2020)(Ghamizi et al., 2023):

- Lack of interpretability and explainability of ML models - The "black box" nature of complex ML models makes it difficult for developers to understand and trust their outputs (Alamin & Uddin, 2021).

- Challenge in defining appropriate test oracles and quality metrics for ML-based systems - The inherent uncertainty and probabilistic nature of ML models do not align well with the deterministic tests and performance thresholds used in traditional software testing

- Difficulty in ensuring privacy and security of data used for ML training - Sensitive data involved in software projects introduces legal and ethical concerns around data handling and model robustness (Liem & Panichella, 2020).

**RQ4**: How do issues like data quality, interpretability of models, and scalability affect the successful implementation of machine learning in software development?

The reviewed literature highlights several data-related and model-centric challenges that impact the real-world deployment of ML in SE:

- Data quality and curation - Software engineering data (e. code, logs, requirements) is often noisy, imbalanced, and incomplete, necessitating significant preprocessing and cleaning.

- Interpretability of ML models - The complex and opaque nature of advanced ML techniques, such as deep learning, makes it challenging for developers to understand how these models arrive at their predictions or recommendations.

- Scalability and performance - Large-scale deployment of ML models for tasks like defect prediction or effort estimation can be computationally intensive and may not meet the performance requirements of agile software development workflows.

**RQ5**: What are the emerging trends and future directions of machine learning in software engineering research and industry applications?

The reviewed sources indicate several promising future directions for ML in software engineering: Combining data-driven machine learning techniques with symbolic AI and knowledge-based systems can enhance the interpretability and trustworthiness of AI systems in software engineering. Researchers and practitioners must prioritize the development of "human-centered AI" solutions that are designed with the end-user in mind, considering factors such as explainability, ethical implications, and user experience. Advanced ML methods like federated learning and differential privacy can help address data quality, privacy, and security concerns in SE applications. Additionally, the integration of ML into software engineering tools and IDEs can make AI-powered capabilities more seamless and accessible to developers in their day-to-day workflows.

In conclusion, the reviewed literature demonstrates the significant potential of machine learning to enhance various software engineering tasks and activities. However, realising the full benefits of ML in software development will require addressing key technical and human-centric challenges around data quality, model interpretability, and ethical AI (Fischer et al., 2020)(Crawford et al., 2023)(Liem & Panichella, 2020). A summary of the literature review findings is provided on Table 1 below.

**Table 1.** A Summary of the literature review findings

| Area of Application | Study & Year | Focus | Key Findings |
|---|---|---|---|
| **Bug Detection** | Michael Pradel et al. (2021) | Use of DL for neural software analysis with DeepBugs | Promising results in detecting potential issue using NLP |

| **Defect Prediction** | Tian et al. (2015) | Predict software defects during code changes using DL | Immediate feedback during code reviews, identification of source code bugs and vulnerabiliti |
| **Automated Bug Fixing** | Allamanis et al. (2021) | Automatically fix build errors with Graph2Diff | Used GNNs to rectify code changes and post commit build errors |
| **Predictive Maintenance** | Hall et al. (2011) | Fault prediction performance in SE | Effective utilisation of decision trees, neural networks, and SVMs to predict software failures |
| **Just-in-Time Defect Prediction** | Tian et al. (2015) | Use of deep belief networks (DBNs) for real-time feedback | Combined code complexity measures and historical defect data for early defect detection |

| | | | Combined code complexity measures and historical defect data for early defect detection |
|---|---|---|---|
| **Maintenance Effort Estimation** | Jindal et al. (2015) | Estimate software maintenance effort using neural networks | Accurate prediction of maintenance effort, facilitating effective resource planning |
| **Bug Localization** | Yousofvand et al. (2023) | Hybrid approach combining DL and model transformation | Enhanced bug localization using CFGs and node classification |
| **Automated Code Generation** | Liao et al. (2010) | Formal and model-driven development methodologies | Critical technique for transforming abstract models into executable code |

| Automated Code Generation | Perez et al. (2021) | Use of pre-trained LLMs like GPT-2 | Leveraged transformer architecture for code generation |
|---|---|---|---|
| **Automated Code Generation** | Sebastián et al. (2020) | NLP and ML techniques for MVC-based applications | Generated source code for MVC-based applications |
| **Code Summarization & Documentation** | Hashemi et al. (2020) | Structured documentation framework for ML projects | Improved quality and usability of ML software documentation |
| **Code Summarization & Documentation** | Barone et al. (2017) | Creation of parallel corpus of Python functions and texts | Created dataset for training ML algorithms for automated documentation |
| **Code Summarization & Documentation** | McBurney and McMillan (2015) | Techniques for automatic Java source code documentation | Static analysis with ML for generating context-providing summaries |

# SECTION 5

**Conclusion**

In conclusion, this paper highlights the significant advancements made in various aspects of software development, including bug detection, software maintenance, automated bug fixing, code generation, and documentation. The studies reviewed demonstrate the potential of machine learning algorithms in improving software quality, reducing maintenance costs, and enhancing the overall efficiency of software development processes. We additionally highlighted the most frequently used algorithms and techniques in software development tasks, such as artificial neural networks (ANN), deep belief networks (DBNs), Natural Language Processing, and Large Language Models (LLMs). This paper emphasises the need for overcoming present research constraints and gaps, such as a lack of comprehensive frameworks that incorporate multiple machine learning techniques for faster source code maintenance. Future studies will focus on ML and DL powered techniques for improvement of software quality maintenance, comprehension and documentation.